\documentclass[
 aip,
 apl,
 reprint,
]{revtex4-1}

\usepackage{graphicx}
\usepackage{upgreek}
\usepackage{epstopdf}
\usepackage{wrapfig}

\begin{document}

\preprint{AIP/123-QED}

\title[Sample title]{Tuning the Dirac point to the Fermi level in the ternary topological insulator (Bi$_{1-x}$Sb$_{x}$)$_{2}$Te$_{3}$}

\author{Jens \surname{Kellner}}
\email{kellner@physik.rwth-aachen.de}
\affiliation{II. Institute of Physics B and JARA-FIT, RWTH Aachen University, Aachen 52074, Germany}
\author{Markus \surname{Eschbach}}
\affiliation{Forschungszentrum J{\"u}lich GmbH, Peter Gr{\"u}nberg Institut (PGI-6), J{\"u}lich 52428, Germany}
\author{J{\"o}rn \surname{Kampmeier}}
\affiliation{Forschungszentrum J{\"u}lich GmbH, Peter Gr{\"u}nberg Institut (PGI-9), J{\"u}lich 52428, Germany}
\author{Martin \surname{Lanius}}
\affiliation{Forschungszentrum J{\"u}lich GmbH, Peter Gr{\"u}nberg Institut (PGI-9), J{\"u}lich 52428, Germany}
\author{Ewa \surname{Mlynczak}}
\affiliation{Forschungszentrum J{\"u}lich GmbH, Peter Gr{\"u}nberg Institut (PGI-6), J{\"u}lich 52428, Germany}
\author{Gregor \surname{Mussler}}
\affiliation{Forschungszentrum J{\"u}lich GmbH, Peter Gr{\"u}nberg Institut (PGI-9), J{\"u}lich 52428, Germany}
\author{Bernhard \surname{Holl{\"a}nder}}
\affiliation{Forschungszentrum J{\"u}lich GmbH, Peter Gr{\"u}nberg Institut (PGI-9), J{\"u}lich 52428, Germany}
\author{Lukasz \surname{Plucinski}}
\affiliation{Forschungszentrum J{\"u}lich GmbH, Peter Gr{\"u}nberg Institut (PGI-6), J{\"u}lich 52428, Germany}
\author{Marcus \surname{Liebmann}}
\affiliation{II. Institute of Physics B and JARA-FIT, RWTH Aachen University, Aachen 52074, Germany}
\author{Detlev \surname{Gr{\"u}tzmacher}}
\affiliation{Forschungszentrum J{\"u}lich GmbH, Peter Gr{\"u}nberg Institut (PGI-9), J{\"u}lich 52428, Germany}
\author{Claus M. \surname{Schneider}}
\affiliation{Forschungszentrum J{\"u}lich GmbH, Peter Gr{\"u}nberg Institut (PGI-6), J{\"u}lich 52428, Germany}
\author{Markus \surname{Morgenstern}}
\affiliation{II. Institute of Physics B and JARA-FIT, RWTH Aachen University, Aachen 52074, Germany}

\date{\today}

\begin{abstract}
In order to stabilize Majorana excitations within vortices of proximity induced topological superconductors, it is mandatory that the Dirac point matches the Fermi level rather exactly, such that the conventionally confined states within the vortex are well separated from the Majorana-type excitation. Here, we show by angle resolved photoelectron spectroscopy that (Bi$_{1-x}$Sb$_{x}$)$_{2}$Te$_{3}$ thin films with $x=0.94$ prepared by molecular beam epitaxy and transferred in ultrahigh vacuum from the molecular beam epitaxy system to the photoemission setup matches this condition. The Dirac point is within 10 meV around the Fermi level and we do not observe any bulk bands intersecting the Fermi level.
\end{abstract}


\maketitle


A topological insulator (TI) is characterized by a bulk energy gap which hosts conducting helical surface states.\cite{M.Konig2007, D.Hsieh2009} These surface states are protected by time reversal symmetry according to a topological Z$_{\mathrm{2}}$ invariant.\cite{Kane2005,Fu2007} A non-trivial  Z$_{\mathrm{2}}$ number, implying helical surface states, is induced, e.g., by spin-orbit coupling which inverts the band order around the gap at some high symmetry points of the Brillouin zone.\cite{Kane,Zhang,Ando} The resulting non-degenerate surface bands are the starting point for the search of Majorana excitations (ME's).\cite{Fu2008} These excitations being its own antiparticle are rigidly pinned to the Fermi level $E_{\mathrm{F}}$.\cite{Alice2012} They are discussed, e.g., as a possible pathway towards topological quantum computation \cite{Nayak,Sau} using its non-Abelian braiding statistics and its intrinsic high degeneracy.\cite{Ivanov,Stern} Several tunneling experiments have found signatures of ME's in solid state systems, such as at the end of a semiconducting nanowire in contact with a superconductor within an axial magnetic field,\cite{V.Mourik,Deng,Das,Rokhinson} at the end of ferromagnetic chains on top of a superconductor with spin-orbit coupling,\cite{S.Nadj-Perge2014} at the surface of possible topological superconductors \cite{Sasaki,Kirzhner,Sasaki2} or at the surface of superconductors on top of topological insulators.\cite{Koren,Koren2} However, none of these experiments are sufficiently controlled to exclude all other possible explanations \cite{Rainis,Kells,Pilkulin,Bagrets,Gibertini,Liu,Lee} for the found peaks at $E_{\rm F}$. Thus, a smoking-gun experiment is still mandatory.\\

Such an experiment could use a TI with a pierced superconductor on top probed by scanning tunneling microscopy (STM).\cite{A.L.Rakhamanov2011} The pierced holes will act as traps for the quantized vortices induced by
magnetic field $B$.
\begin{wrapfigure}{o}{0.4\columnwidth}
\includegraphics[width=0.4\columnwidth]{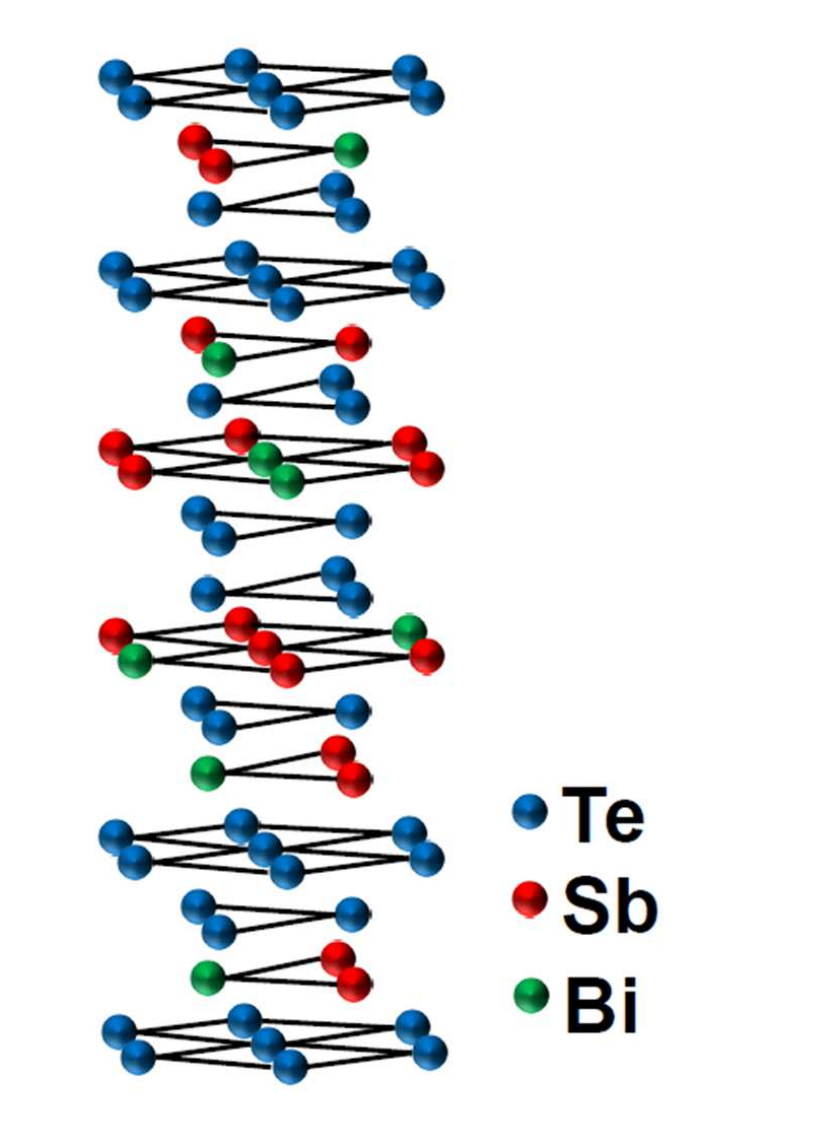}
\caption{Atomic structure of (Bi$_{1-x}$Sb$_{x}$)$_{2}$Te$_{3}$ (tetradymite-type crystal)
with colors of different atoms marked.
\label{fig:Structure}}
\end{wrapfigure}
 If the number of vortices within a hole is odd, it will provide a ME at $E_{\rm F}$\cite{Bolech,Tewari} with predicted spatial distribution, \cite{A.L.Rakhamanov2011} while, if even, at slightly higher or slightly lower $B$, the  ME will disappear in favor of conventional excitations away from $E_{\rm F}$ \cite{Alice2012}. In order to realize such an experiment, the ME's have to be energetically well separated from conventional confined states within the topological vortex, i.e. the density of states of the Dirac cone of the TI has to be sufficiently low.\cite{J.-P.Xu2014} If the hole size is about the coherence length of the superconductor, one can estimate the energetic distance of the ME to the first excited state by $E_{\mathrm{1}}-E_{\mathrm{F}} \simeq \Delta^{2}/\sqrt{\Delta^2+(E_{\rm F}-E_{\rm D})^2}$ with $E_{\rm D}$ being the Dirac point energy and $\Delta$ the superconducting gap.\cite{J.DSau2010,A.L.Rakhamanov2011}
For a typical energy resolution of 0.1 meV at $T = 0.3$ K \cite{Wiebe} and $\Delta\simeq 1-2$ meV, this implies $|E_{\rm D}-E_{\rm F}|\leq 10$ meV. Additionally, the Dirac point has to be energetically remote from any bulk states.\\
This cannot be achieved with binary TI's as, e.g., Bi$_{\mathrm{2}}$Te$_{\mathrm{3}}$, Sb$_{\mathrm{2}}$Te$_{\mathrm{3}}$ or Bi$_{\mathrm{2}}$Se$_{\mathrm{3}}$.\cite{H.Zhang12009} For Bi$_{\mathrm{2}}$Te$_{\mathrm{3}}$, $E_{\mathrm{D}}$ is buried in the bulk valence band (BVB) and $E_{\mathrm{F}}$ is located in the bulk conduction band (BCB), whereas for Sb$_{\mathrm{2}}$Te$_{\mathrm{3}}$, $E_{\mathrm{D}}$ is in the bulk energy gap and $E_{\mathrm{F}}$ is located in the BVB. \cite{H.Zhang12009}\\

Mixing these two compounds leads to the ternary system (Bi$_{1-x}$Sb$_{x}$)$_{2}$Te$_{3}$, which exhibits the same tetradymite structure as Bi$_{\mathrm{2}}$Te$_{\mathrm{3}}$ and Sb$_{\mathrm{2}}$Te$_{\mathrm{3}}$ with a mix of Sb and Bi atoms in one of the layers (Fig. \ref{fig:Structure}). Previous angle resolved photoelectron spectroscopy (ARPES) and transport experiments have shown that this ternary alloy exhibits $E_{\rm F}-E_{\rm D}\simeq 70$ meV at  $x\approx0.94$ and $E_{\rm F}-E_{\rm D}\simeq -30$ meV at $x\approx0.96$.\cite{J.Zhang2011}  This indicates that the favorable condition of $E_{\rm F}\simeq E_{\rm D}$ is possible at an $x$ in between. Moreover, the transport data are compatible with a charge carrier concentration originating from the Dirac cone only, such that one can anticipate the absence of bulk bands at $E_{\rm D}$, albeit the authors did not study the bulk bands in detail by ARPES.\cite{J.Zhang2011} \\
Here, we investigate thin films of (Bi$_{1-x}$Sb$_{x}$)$_{2}$Te$_{3}$ on Si(111) by ARPES, after an ultra-high vacuum (UHV) transfer from the molecular beam epitaxy (MBE) chamber to the ARPES setup. We established a sufficient accuracy of $|E_{\mathrm{D}}$ - $E_{\mathrm{F}}|$ for the selection of samples for the mentioned Majorana experiments, which is applied to several stoichiometries  ($x=0.48$, $0.82$, $0.93$ and $0.94$). For $x=0.94$, we show that  $|E_{\mathrm{D}}$ - $E_{\mathrm{F}}|\leq 2$ meV $\pm 7$ meV, which has to be contrasted with the previous precision of about 50 meV of [38,40]. Furthermore, we confirm the absence of bulk states at  $E_{\mathrm{F}}$, at least, for photon energies $h\nu=8.44$ eV and 21.2 eV. A similar condition has been achieved previously for the quaternary compound BiSbTeSe$_{2}$,\cite{T.Arakane2012, Y.Xu2014} but not for any ternary compound. Such a ternary might be more easy to handle in terms of stoichiometry and interface chemistry, which is probably also favorable for transport and spin transport experiments requiring $E_{\rm D}$ close to $E_{\rm F}$.\cite{J.Zhang2011, R.Yoshimi2014, J.Tang2014, D.Kong2011} Here, we only show the data for $x=0.94$, where $E_{\mathrm{D}}$ is closest to $E_{\mathrm{F}}$.\\

\begin{figure}
\includegraphics[width=1\linewidth]{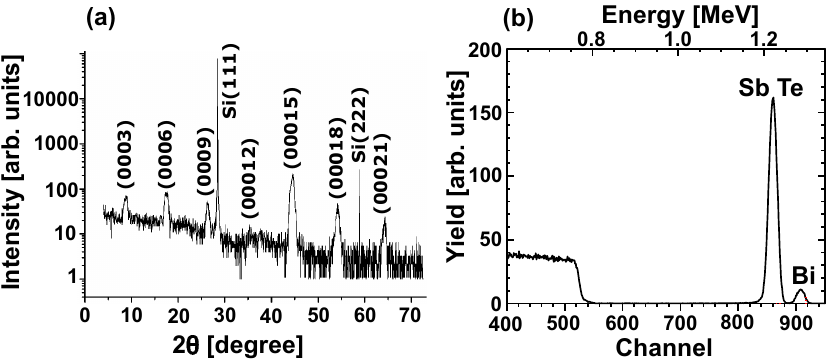}
\caption{(a) XRD $2\Theta/\Theta$ scan; (000$ l$) reflections of the grown thin film and Si(111) substrate reflections are labeled; (b) RBS measurement to determine the stoichiometry; Sb Te peak and Bi peak are labeled. \label{fig:XRDRBS}}
 \end{figure}
\begin{figure}
\includegraphics[width=1\linewidth]{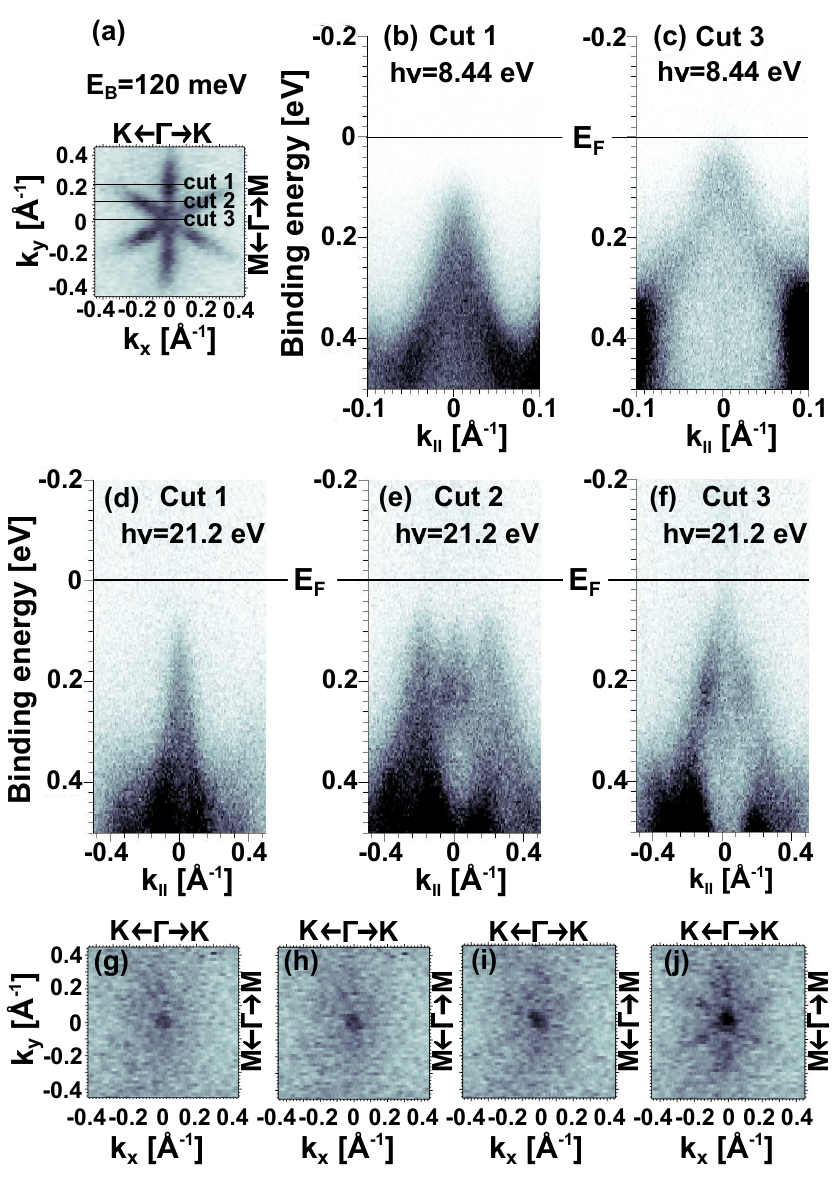}
 \caption{ARPES data probing the bulk valence band (BVB) (MBE growth, in-situ transfer):
(a) CEC at $E_{\rm B}=120$ meV showing the star shaped maximum area of the BVB, where $k_{x}$ and $k_{y}$ are in the $\Gamma$ - K and $\Gamma$ - M directions, respectively; $h\nu=8.44$ eV; cuts mark the $k_{\rm ||}$ directions of the plots in (b)$-$(f) as labeled; (b)$-$(f) ARPES data in $E(k_{||})$ representation along the cuts marked in (a) with photon energies indicated; the solid black line $E_{\rm F}$ is determined on polycrystalline Cu; (g)-(j) CEC's at (g) $E_{\mathrm{B}}$ =0 meV, (h) 10 meV, (i) 20 meV, (j) 40 meV, $h\nu=21.2$ eV.
\label{fig:EDC-VB}}
\end{figure}

Thin films of (Bi$_{1-x}$Sb$_{x}$)$_{2}$Te$_{3}$ were grown by MBE on highly doped Si(111) substrates, prepared \textit{in-situ} by flash annealing in UHV. Knudsen effusion cells were used for the evaporation at a substrate temperature of $T=275^{\circ}$C. X-ray diffraction (XRD) and reflectivity (XRR) measurements were performed \textit{ex-situ}. The consecutive order of reflections correspond to (000$l$) reflections of hexagonal Bi$_{2}$Te$_{3}$, evidencing that the (Bi,Sb)$_{2}$Te$_{3}$ film is of single crystal nature (Fig. \ref{fig:XRDRBS}(a)). XRR reveals a film thickness of $23\pm1$ nm. Rutherford backscattering spectrometry (RBS) determies the Bi yield to $2.10\times 10^{15}$ atoms/cm$^{2}$ and the merged Sb and Te yield to $8.8\times 10^{16}$ atoms/cm$^{2}$ (Fig. \ref{fig:XRDRBS}(b)). Combined with the XRD measurement, the stoichiometry is calculated to be $x=0.94\pm0.01$ (for a more detailed discussion, see supplement \cite{supplement}).\\
\begin{figure}
\includegraphics[width=1\linewidth]{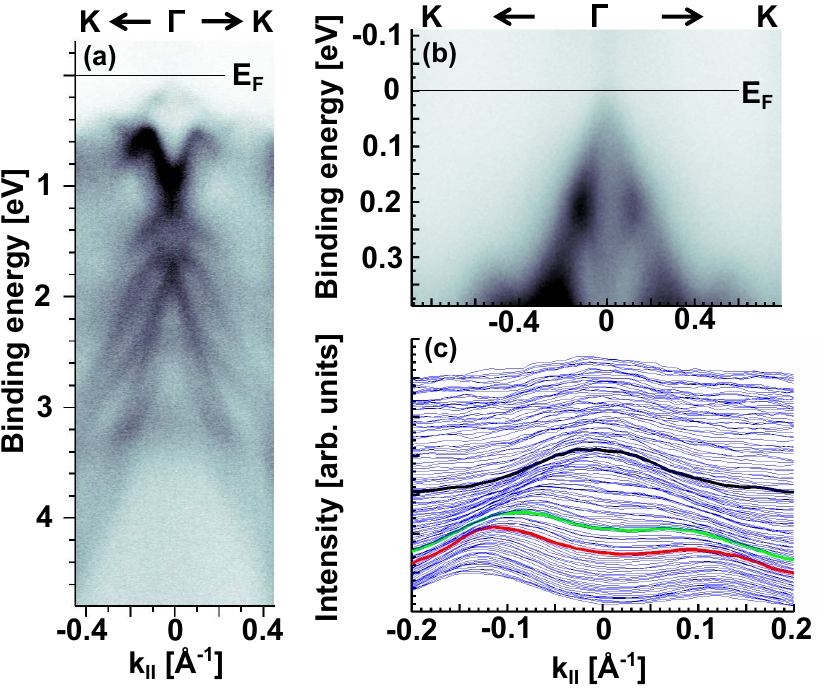}
\caption{ARPES data of (Bi$_{0.06}$Sb$_{0.94}$)$_{2}$Te$_{3}$, $h\nu=21.2$ eV, T= 200 K (MBE growth, in-situ transfer):
(a),(b) $E(k_{||})$ dispersion along $\Gamma$ - K at different resolution (black: high intensity, white: low intensity), Fermi level $E_{\rm F}$ as determined on polycrystalline Cu is marked as a solid black line; (c) Momentum distribution curves (MDCs) between binding energies $E_{\mathrm{B}}$ = 0.240 eV and $E_{\mathrm{B}}$  = -0.110 eV, continuously offset for clarity; red and black MDC's are at $E_{\mathrm{B}}=0.170$ eV and $E_{\mathrm{B}}=0.045$ eV, respectively; green MDC corresponds to Fig. \ref{fig:MDC}(a) \label{fig:ARPES_full_spectrum}}
 \end{figure}
The samples were transferred \textit{in-situ} from the MBE system to the ARPES system with a vacuum shuttle at pressure $p=5\times10^{-10}$ mbar. This prevents oxidation and surface contamination of the sample. The ARPES spectra were recorded at $T = 50$ K with a MBS A1 hemispherical analyzer providing an energy resolution of $10$ meV and an angular resolution of $0.1{^\circ}$. Two different $h\nu$ are used provided by the non-monochromatized He I$\alpha$ resonance ($h\nu=21.22$ eV) and monochromatized light from a microwave-driven Xenon source ($h\nu=8.44$ eV). For both $h\nu$, $E_{\mathrm{F}}$ has been determined independently on polycrystalline Cu with an accuracy of 7 meV. There are no indications of time-dependent band bending within 2 days in contrast to similar experiments on Bi$_{2}$Se$_{3}$,\cite{King, Benia} i.e. the band bending is below 7 meV. Photovoltage effects on (Bi$_{0.06}$Sb$_{0.94}$)$_{2}$Te$_{3}$ have been checked to be negligible by the temperature dependence of the spectra between 50 K and 300 K.
The  directions in momentum ($k_{||}$) space of the ARPES spectra are deduced by comparison with ab-inito calculations for Sb$_{2}$Te$_{3}$(0001),\cite{Pauly} i.e., the lips of the star-like structure in Fig. \ref{fig:EDC-VB}(a) point to $\Gamma$-M direction. 

\begin{figure}
\includegraphics[width=1\linewidth]{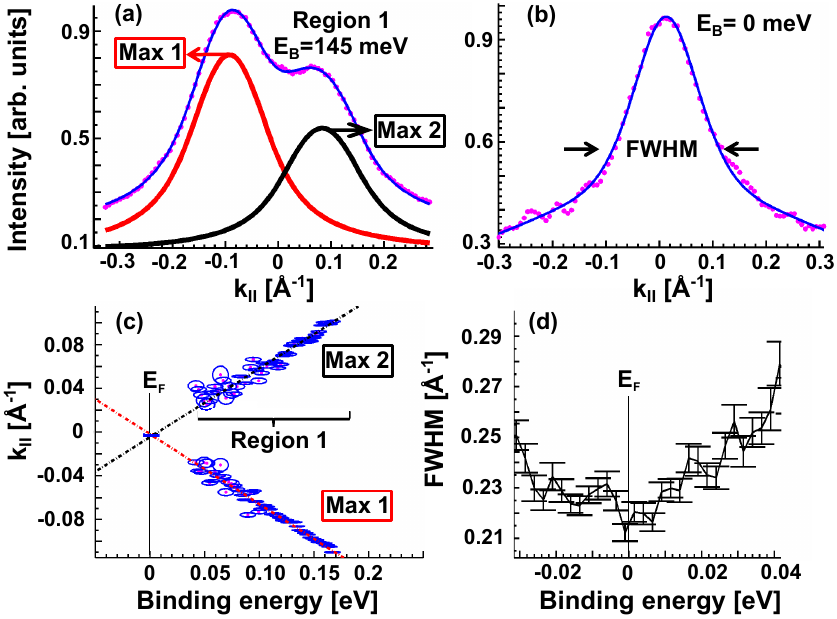}
\caption{
(a) MDC's (purple dots) at indicated $E_{\mathrm{B}}$ with two Voigt fit curves (black, red) with indicated peak positions (Max 1, Max 2) and resulting fit (blue curve); (b) same as (a) with only one Voigt fit curve (blue) and marked FWHM; (c) peak positions deduced from fitting as shown in (a), as function of $E_{\mathrm{B}}$; only region 1 is used for fits and the linear regression (black, red dashed line) to determine  $E_{\mathrm{D}}$; error bars (blue ellipses) result from the fitting procedure; vertical error bar (blue) at $E_{\mathrm{D}}$ is the error on the intersection point; (d) FWHM's of the MDC peak close to $E_{\rm F}$ determined as marked in (b). \label{fig:MDC}}
\end{figure}

The measured band structure in $\Gamma$-K direction is shown in Fig. \ref{fig:ARPES_full_spectrum}(a)$-$(b). The characteristic linear band dispersion of the Dirac type topological surface state (TSS) is visible (Fig. \ref{fig:ARPES_full_spectrum}(b)) having the Dirac point close to $E_{\rm F}$. This can also be deduced from the momentum distribution curves (MDC's) in Fig. \ref{fig:ARPES_full_spectrum}(c).
To evaluate the position of $E_{\mathrm{D}}$  more precisely, two Lorentzian peaks convoluted with a Gaussian function (Voigt function \cite{Valla,Valla2,Valla3}) were fitted to the MDC's of the TSS  (Fig. \ref{fig:MDC}(a)) for binding energies $E_{\mathrm{B}}=0.170-0.045$ eV (region 1, 9 fit parameters per MDC, see supplement \cite{supplement}). The
$k_{||}$ positions of the resulting peak maxima are plotted in Fig. \ref{fig:MDC}(c) exhibiting an average error of $\Delta \overline{k}=\pm1.5\times10^{-3}$ ${\rm \AA}^{-1}$. For $E_{\mathrm{B}}=-0.030-0.045$ eV, the two maxima are too close to be separately visible in the MDC's resulting in much larger errors up to $\Delta \overline{k}=\pm1.5\times10^{-2}$ ${\rm \AA}^{-1}$. To determine the position of  $E_{\mathrm{D}}$, a linear fit of the data for $E_{\mathrm{B}}=0.170-0.045$ eV (region 1 of Fig. \ref{fig:MDC}(c)) is extrapolated to $E_{\mathrm{F}}$. The linear regression, which includes the $k_{||}$ errors from the Voigt fits and the energy errors resulting mostly from the uncertainty of $E_{\mathrm{F}}$, reveals a Dirac velocity for the TSS at $-k_{\rm ||}$ of $v_{\mathrm{D}}=(3.8\pm0.2)\times10^{5}\,\mathrm{\frac{m}{s}}$ and at $k_{\rm ||}$ of $v_{\mathrm{D}}=(3.9\pm0.2)\times10^{5}\,\mathrm{\frac{m}{s}}$. The intersection of the two TSS is at  $E_{\mathrm{B}}=2\pm 7$ meV.

Another way to determine $E_{\rm D}$ is given by the full width at half maximum (FWHM) of a single Voigt peak to the MDC's close to $E_{\rm F}$ (Fig. \ref{fig:MDC}(b)). Due to thermal excitations, this peak can also be recorded above $E_{\rm F}$. The FWHM's of the peaks are plotted in Fig. \ref{fig:MDC}(d) showing a minimum at $E_{\rm B}=0\pm 10$ meV. Thus, (Bi$_{0.06}$Sb$_{0.94}$)$_{2}$Te$_{3}$ exhibits $E_{\mathrm{D}}=E_{\mathrm{F}}$ with a precision better than 10 meV. This is exactly the precondition for the search of ME's by STM as described above, albeit the superconductor on top might shift $E_{\rm D}$ again.\cite{deJong}

Another requirement is, that the BVB is further away from $E_{\mathrm{F}}$ than the induced superconducting gap. In order to probe this, we plot ARPES data through the 
maximum of the BVB in $k_{\rm ||}$ space (along cut 1 and cut 2 in Fig. \ref{fig:EDC-VB}(a)) and the TSS (along cut 3 in Fig. \ref{fig:EDC-VB}(a)) at two different $k_z$, i.e. two different $h\nu$ (Fig. \ref{fig:EDC-VB}(b)-(f)). The valence band maximum is found at $E_{\mathrm{B}}\simeq 60$ meV for both $h\nu$, indicating little dispersion along $k_z$ as expected for the layered material. The absence of the BVB at $E_{\rm F}$ is corroborated by the constant energy cuts (CEC's) at $h\nu =21.2$ eV (Fig. \ref{fig:EDC-VB}(g)-(j)) which exhibit the onset of a BVB band structure at $E_{\rm B} \simeq 40$ meV.\\

In summary, we synthesized a ternary topological insulator (Bi$_{0.06}$Sb$_{0.94}$)$_{2}$Te$_{3}$, which exhibits  $E_{\mathrm{D}}-E_{\mathrm{F}} =2 \pm 7$ meV and does not show any indications of the bulk valence band close to $E_{\rm F}$. This is favorable for both, transport measurements, in particular, probing spin-transport\cite{L-Barreto2014} and a good starting point for the search of Majorana quasiparticles by STM.\cite{Alice2012}


\begin{acknowledgments}
We gratefully acknowledge  financial support by the German science foundation via Mo 858/13-1 and the BMBF (05K13PA4).
\end{acknowledgments}

\end{document}


\titleformat{\section}{\normalfont\Large\bfseries}{\thesection}{1em}{}

\section*{Supplement: Tuning the Dirac point to the Fermi level in the ternary topological insulator (Bi$_{1-x}$Sb$_{x}$)$_{2}$Te$_{3}$}

\subsection{Fitting routine} 

\begin{figure}[h]
\includegraphics[width=1\linewidth]{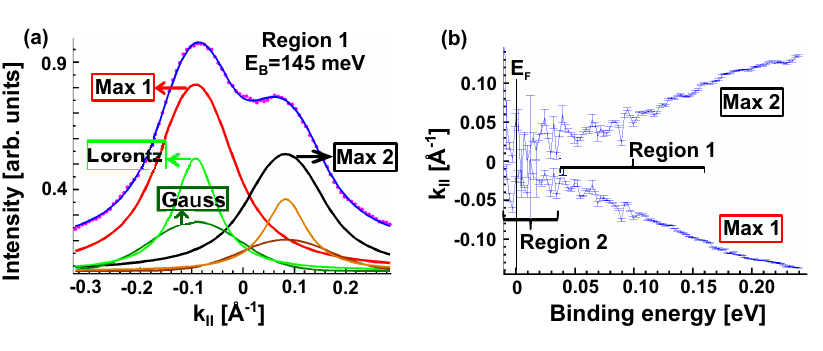}
\caption{(a) MDC's (purple dots) at indicated $E_{\mathrm{B}}$ with two Voigt fit curves (black, red) and resulting fit (blue curve); Lorentz and Gauss contributions for the first peak (light green, dark green curves) and the second peak (light brown, dark brown curves) are added; (b) peak positions deduced from fitting as shown in (a) for $E_{\mathrm{B}}=-0.01$ eV $-$ $0.24$ eV with different regions as described in the text marked. \label{fig:ARPES_supplement}}
 \end{figure}

To quantitatively determine the Dirac point position, the MDC's of the TSS are fitted, firstly, with two Voigt functions. A Voigt function $V_{i}(x)$ is a convolution of a Lorentzian and a Gaussian function, such that: 
\[
\sum_{i=1}^{2}V_{i}(x)=\sum_{i=1}^{2}[y+\frac{2A_{i}}{\pi}\times\frac{w_{\mathrm{L}}}{4(x-x_{\mathrm{c}_{i}}+w_{\mathrm{L}}^{2}))^{2}}+(1-\mu_{i})\times\frac{\sqrt{4\times \mathrm{ln}(2)}}{\sqrt{\pi}\times w_{\mathrm{G}}}\times \mathrm{exp}(-\frac{4\times \mathrm{ln}(2)}{w_{\mathrm{G}}^{2}}\times(x-x_{\mathrm{c}_{i}})^{2})]
\]

where $y$ ist the offset, $A_{i}$ the amplitudes, $x_{\mathrm{c}_{i}}$ the centers of the Voigt functions, $\mu_{i}$ the profile shape factors and $w_{\mathrm{L}}$, $w_{\mathrm{G}}$ the widths of the Lorentzian
and the Gaussian contribution, respectively. The two Voigt functions were adjusted to the MDC's  (Fig. S\ref{fig:ARPES_supplement}(a)) within the energy range $E_{\mathrm{B}}=-0.01$ eV $-$ $0.24$ eV.  In total, the fit has 9 free parameters. The fit is adjusted by a Levenberg-Marquardt algorithm and the standard deviation is read out for every parameter. Note that the widths $w_{\mathrm{L}}$ and $w_{\mathrm{G}}$ are not assigned with an index $i$, thus, the widths are the same for the two peaks at $-k_{\rm ||}$ and $k_{\rm ||}$.\\
The resulting total amplitude, as well as the Lorentzian and Gaussian amplitudes are higher for all $-k_{\rm ||}$. This is most likely due to the geometrical setup of the ARPES system, i.e., matrix element effects induce the difference in intensity.\\ 
The fit results for the peak positions are shown in Fig. S\ref{fig:ARPES_supplement}(b) including error bars. In the region labelled region 2, the error bars are quite large, since the MDC exhibits only one maximum. Moreover, the dispersion deviates from linearity for $E_{\mathrm{B}}>0.170$ eV. Consequently, we used only the energy region $E_{\mathrm{B}}=0.170$ eV $-$ $0.045$ eV (region 1) for the determination of $E_{\mathrm{D}}$ as shown in Fig. 4(c) of the main text.

\newpage

\subsection{RBS and XRD measurement } 

\begin{figure}[h]
\includegraphics[width=1\linewidth]{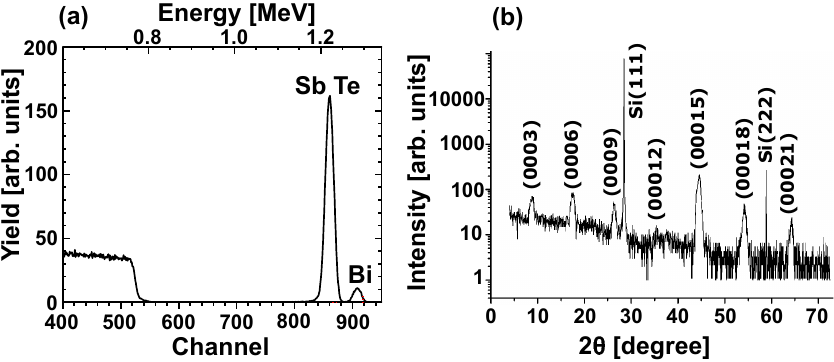}
\caption{(a) RBS measurement to determine the stoichiometry; Sb Te peak and Bi peak are labeled; (b) XRD $2\Theta/\Theta$ scan; (000$ l$) reflections of the grown thin film and Si(111) substrate reflections are labeled. \label{fig:stoichiometry_supplement}}
 \end{figure}

Rutherford backscattering spectrometry (RBS) using $1.4$ MeV He$^{+}$ ions at a scattering angle of $170$ degrees was applied in order to determine the stoichiometry of the film. Fig. S\ref{fig:stoichiometry_supplement}(a) shows a typical RBS spectrum. The absolute areal coverage was determined by integration of the two peaks. The peak of Bi is clearly separated and yields $2.10\times 10^{15}$ atoms/cm$^{2}$.  Due to the low difference of the atomic mass of Sb and Te these two peaks cannot be resolved separately. Therefore, only the sum of Sb and Te atoms of $8.8\times 10^{16}$ atoms/cm$^{2}$ is determined. The experimental accuracy of the absolute values is estimated to be $\pm 2 \%$.\\
X-ray diffraction (XRD) measurements were accomplished employing a Bruker D8 high-resolution diffractometer. The XRD scans were carried out in a symmetric $2\Theta/\Theta$ configuration. Figure S\ref{fig:stoichiometry_supplement}(b) depicts the XRD curve of the investigated (Bi,Sb)$_{2}$Te$_{3}$ sample. Numerous peaks originating from the (Bi,Sb)$_{2}$Te$_{3}$ are seen, which can be assigned to reflections in the (000$l$) direction with an accuracy of $\pm 0.5^{\circ}$. The consecutive order of reflections correspond to (000$l$) reflections of hexagonal Bi$_{2}$Te$_{3}$, evidencing that the (Bi,Sb)$_{2}$Te$_{3}$ film has separate BiSb and Te layers according to the stacking sequence shown in Fig. 1 of the main text. If the deviation from the crystal stoichiometry (Bi,Sb)$_{2}$Te$_{3}$ would be more than $\pm 1 \%$, the resulting distortion of the crystal would shift the XRD peaks by more than  $\pm 1^{\circ}$. Thus the accuracy of the stoichiometry is estimated to be  $\pm 1 \%$.

Combining the RBS and XRD results, the Sb concentration can be computed to $x=94\%\pm1\%$ within the formula  (Bi$_{1-x}$Sb$_{x}$)$_{2}$Te$_{3}$.